\documentclass{aa} 
\usepackage[dvipsnames]{xcolor} 
\usepackage{graphicx, subfigure, amsmath}
\usepackage[breaklinks=true]{hyperref} 
\hypersetup{colorlinks=true,citecolor=Blue}

\begin{document}

   \title{Resolved millimeter-dust continuum cavity around the very low mass young star CIDA\,1}


   \author{
   Paola Pinilla\inst{1, 2},
   Antonella Natta\inst{3,4},
   Carlo F. Manara\inst{5},
   Luca Ricci\inst{6, 7},
   Aleks Scholz\inst{8}, and
   Leonardo Testi\inst{3, 5}
          }

   \institute{Department of Astronomy/Steward Observatory, The University of Arizona, 933 North Cherry Avenue, Tucson, AZ 85721, USA,      \email{pinilla@email.arizona.edu}
    \and Hubble Fellow
    \and INAF-Arcetri, Largo E. Fermi 5, I-50125 Firenze
    \and Dublin Institute for Advanced Studies, School of Cosmic Physics, 31 Fitzwilliam Place, Dublin 2, Ireland
    \and European Southern Observatory, Karl-Schwarzschild-Str. 2, D85748 Garching, Germany
    \and Department of Physics and Astronomy, California State University Northridge, 18111 Nordhoff St, Northridge, CA 91130, USA
   \and Jet Propulsion Laboratory, California Institute of Technology, 4800 Oak Grove Drive, Pasadena, CA, 91109, USA
    \and SUPA, School of Physics \& Astronomy, University of St. Andrews, North Haugh, St. Andrews, KY16 9SS, UK       
     }

   \date{}

 
  \abstract
   {Transition disks (TDs) are circumstellar disks with inner regions highly depleted in dust. TDs are observed in a small fraction of disk-bearing objects at ages of 1-10\,Myr. They are important laboratories to study evolutionary effects in disks, from  photoevaporation to planet-disk interactions.}  
   {We report the discovery of a large inner dust-empty region in the disk around the very low mass star CIDA\,1 (M$_{\star} \sim 0.1-0.2$ M$_{\odot}$).}
   {We used ALMA continuum observations at 887\,$\mu$m, which provide a spatial resolution of $0.\arcsec21\times0.\arcsec12$ ($\sim$15$\times$8\,au in radius at 140\,pc).} 
   {The data show a dusty ring with a clear cavity of radius $\sim$20\,au, the typical characteristic of a TD. The emission in the ring is well described by a narrow Gaussian profile. The dust mass in the disk is $\sim$17\,M$_{\oplus}$. CIDA\,1 is one of the lowest mass stars with a clearly detected millimeter cavity. When compared to objects of similar stellar mass, it has a relatively massive dusty disk (less than $\sim5$\% of Taurus Class II disks in Taurus have a ratio of $M_{\rm{disk}}/M_{\star}$ larger than  CIDA\,1) and a very high mass accretion rate (CIDA\,1 is a disk with one of the lowest values of  $M_{\rm{disk}}/\dot M$ ever observed). In light of these unusual parameters, we discuss a number of possible mechanisms that can be responsible for the formation of the dust cavity (e.g., photoevaporation, dead zones, embedded planets, close binary). We find that an embedded planet of a Saturn mass or a close binary are the most likely possibilities.}
   {}

   \keywords{accretion, accretion disk -- circumstellar matter --stars: premain-sequence-protoplanetary disk--planet formation}

   \titlerunning{Resolved cavity around the VLM young star CIDA\,1}
    \authorrunning{P.~Pinilla et al.}
    \maketitle

%
\section{Introduction}			\label{sect:intro}

Protoplanetary disks around very low mass (VLM) stars and brown dwarfs (BDs) are excellent laboratories to investigate planet formation in extreme conditions, such as very low disk mass and temperature. Recent exoplanet discoveries around VLM stars provide exciting scenarios where several Earth-like planets are tightly packed in the habitable zone \citep[e.g., TRAPPIST-1, ][]{gillon2017}. However, it is still unclear how planets overall form around VLM stars. 

In general, the first steps of planet formation involve the dynamics and growth of small (submicron-sized) particles   to planetesimals \citep[e.g.,][]{armitage2010}. These processes are regulated by the interaction of the dust with the gas. Due to the gas drag, dust particles experience a fast radial drift toward the star. This radial drift can be a barrier for growth since it is expected to happen in very short timescales before particles grow to planetesimals \citep{brauer2008}. 

In disks around VLM stars and BDs, the radial drift barrier is more difficult to overcome, since the drift velocities are expected to be higher than around T-Tauri or Herbig stars \citep{pinilla2013, zhu2017}. This is because the dust drift velocity depends on the the difference between the orbital gas velocity and the Keplerian speed, and as a result $v_{\rm{drift}}\propto1/\sqrt{M_\star}$ \citep[see mathematical derivation in Sect. 2.2 in][]{pinilla2013}. It is, however, evident that disks around VLM stars and BDs host millimeter sized-particles \citep{Ricci2014, bayo2017, pinilla2017b} and that somehow the radial drift of the particles is reduced or completely suppressed in these disks.

A potential possibility for reducing the radial drift of grains is the existence of local pressure bumps where grains accumulate and can continue growing to planetesimals \citep{whipple1972, weidenschilling1977}. This idea has had a growing interest due to the substructures that have been resolved in the last years from multiwavelength observations of protoplanetary disks, which are suggested to originate from particle trapping in pressure bumps \citep[e.g.,][]{alma2015, andrews2016, isella2016, ginski2016, plas2016, boekel2017, fedele2017, cieza2017}.
In the case of TRAPPIST-1, recent models invoke the rapid formation of planetesimals at the snow line in combination with streaming instabilities \citep[e.g.,][]{ormel2017}.

 In this context, of special interest  is the structures observed in transition disks (TDs), which are disks with empty dust cavities, and which were originally identified by a lack of near and mid-infrared emission in the spectral energy distribution \citep[SED,][]{strom1989}. Recent observations with ALMA have reveal that some disks that seem to be full disks from the SED, may host a large millimeter cavity \citep[e.g., RY\,Lup, Sz118][]{ansdell2016,vandermarel2018}. One of the most frequent explanations for the origin of TDs structures is the existence of a massive planet filtering dust particles and trapping mainly the millimeter sized particles in the outer pressure bump formed at the edge of the planetary gap \citep[e.g.,][]{rice2006, zhu2011, pinilla2012}.  Most of the resolved millimeter cavities have been observed around Herbig Ae/Be stars and early type (F, G, K) T-Tauri stars \citep[e.g.,][]{marel2016,vandermarel2018}. Disk masses scale with the stellar mass \citep{andrews2013}, hence it may be easier to detect TDs around Herbig Ae/Be and T-Tauri stars than VLM stars, because they have higher amount of material to form massive planets. Nevertheless, this can also be the result of an observational bias since disks around BDs and VLM stars are are fainter and smaller and thus more difficult to detect.

In this paper we report a resolved millimeter-dust continuum cavity around a VLM young star known as CIDA\,1 from ALMA observations. In the next Sect~\ref{sec:target} we describe the observed properties of CIDA\,1. Section~\ref{sec:observations} summarizes our ALMA observations and data reduction. Sections~\ref{sec:results} and \ref{sec:analysis} present the summary of the results from the observations and the data analysis respectively. Finally, in Sects.~\ref{sec:discussion} and \ref{sec:conclusions} we discuss our results in the context of the potential origin of the observed cavity, and the conclusions of this work.  

\section {CIDA\,1}			\label{sec:target}

\begin{figure}
 \centering
  	\includegraphics[width=\columnwidth]{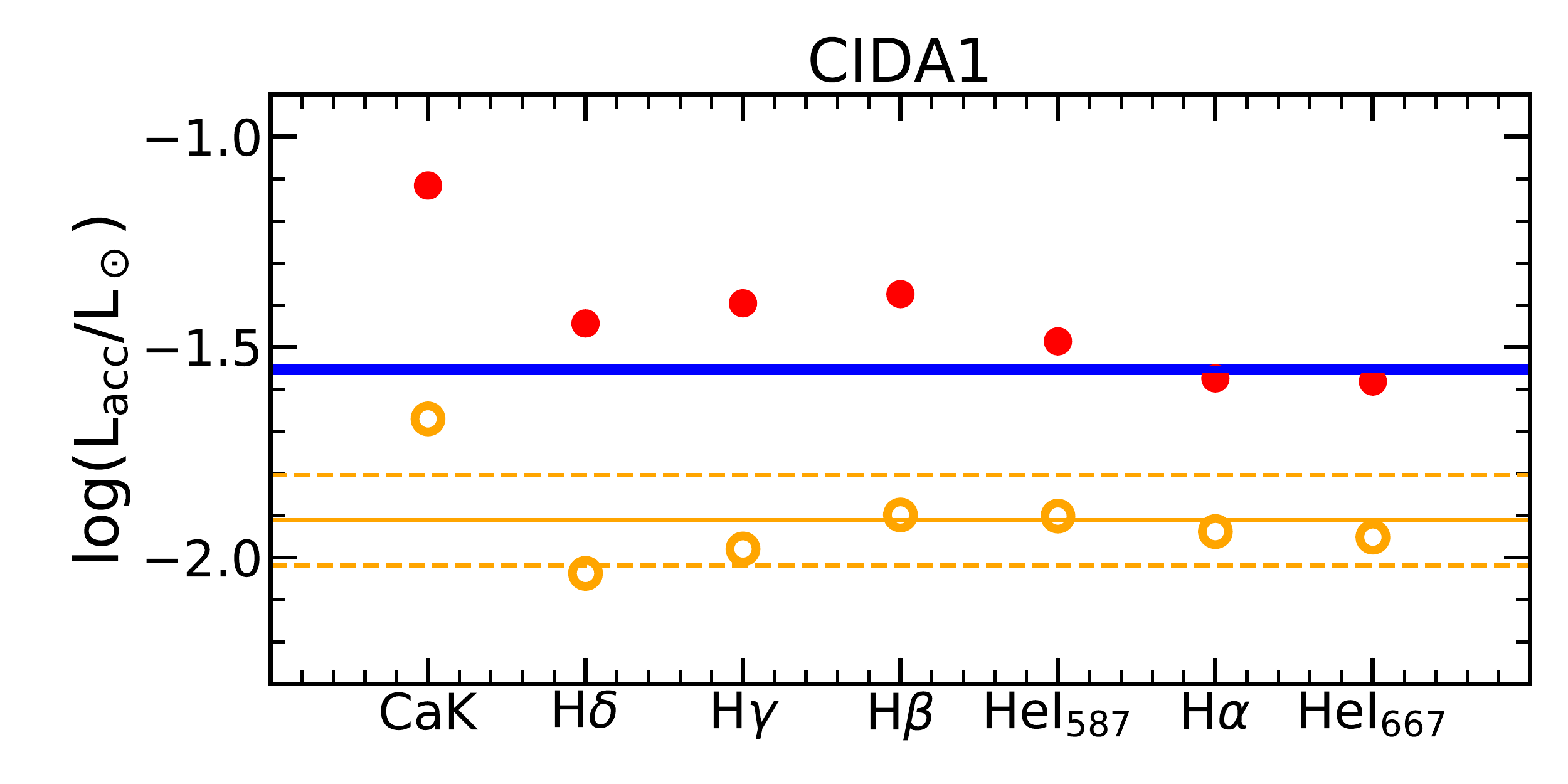}
   \caption{Accretion luminosity obtained from the luminosity of different lines using the relations of \citet{Alcala2014}. Red points are for values obtained assuming $A_V$=3 mag, open yellow points assuming $A_V$=2 mag. The blue line is the value of accretion luminosity derived from Balmer continuum excess modeling assuming $A_V$=3 mag. A smaller spread of values of accretion luminosity is present when $A_V$ = 2 mag. }
   \label{lacc_lines_cida1}
\end{figure}

CIDA\,1 (J04141760+2806096) is a low mass Class II T-Tauri star in Taurus \citep{briceno1993}. Its spectral type has been estimated to be between M4.5 \citep{Herczeg2014} and M5.5 \citep{WhiteBasri2003}, with a non-negligible extinction of $A_V \sim 3$ mag. The luminosity quoted in the literature is $\sim 0.2$ L$_\odot$ \citep{Herczeg2008,Herczeg2014}, with a corresponding mass in the range 0.1--0.2 M$_\odot$, depending on the adopted evolutionary tracks. The effective temperature is estimated to be 3000-3100\,K. CIDA\,1 is a strong accretor for its stellar mass. \cite{Herczeg2008} estimate $\dot M\sim 1.2\times 10^{-8}$ M$_\odot$/yr for $A_V \sim 3$ mag.

This value of extinction reported in the literature for CIDA\,1 is rather uncertain \citep{Herczeg2008}. 
We measure the extinction toward CIDA\,1 using an indirect, but powerful method. We assume that the accretion luminosity L$_{\rm{acc}}$ can be derived with good accuracy from the luminosity of several different emission lines, using the relations derived by \cite{Alcala2014}. If properly corrected for extinction, all the lines should give the same result \citep[e.g.,][]{Alcala2014, Rigliaco2012}. We use the fluxes of the H$\alpha$, H$\beta$, H$\gamma$, H$\delta$, CaK $\lambda$393.4 nm, He $\lambda$587.56 nm, and He $\lambda$667.82 nm measured by \citet{Herczeg2008} to determine $L_{\rm{acc}}$ assuming different values of $A_V$ from 0 to 3 mag. The standard deviation of the values of log$L_{\rm{acc}}$ determined from the various lines at different values of $A_V$ presents a minimum at $A_V$ = 2 mag, as shown in Fig.~\ref{lacc_lines_cida1}. In this figure, the red dots are used for the values of $L_{\rm{acc}}$ for $A_V=3$ mag, and the blue line the value derived by \cite{Herczeg2008} from the UV excess continuum emission, corrected for the same extinction. The yellow dots are the line-derived values for $A_V=2$ mag. We thus assume that $A_V$ is 2 mag.

Using this extinction in combination with the 2MASS J-band magnitude and a distance of 140\,pc (typical for young stars in the Taurus association) gives an absolute J-band magnitude of $M_J = 5.4$. Comparing with the \citet{Baraffe1998} 1\,Myr evolutionary track yields L$_\star$=$8\times10^{-2}$L$_\odot$ and M$_\star$=0.1\,M$_\odot$$\pm 0.1$\,dex. From the hydrogen recombination line fluxes, we derive a value of the accretion luminosity of $1.2\times10^{-2}$\,L$_\odot$. The corresponding mass accretion rate is $4.0\times 10^{-9}$ M$_\odot$/yr  $\pm 0.45$\,dex \citep[the uncertainty is based on the cumulative relative uncertainty calculated in][]{Alcala2014}. Comparing the effective temperature with the same evolutionary track gives a very similar estimate for mass and luminosity. Switching to the more recent \citet{Baraffe2015} isochrone for 1\,Myr again yields consistent results for the stellar parameters.

CIDA\,1 hosts a circumstellar disk, which was first detected in the millimeter domain by \citet{Schaefer2009}. ALMA Cycle 0 observations showed evidence of a large dusty disk \citep{Ricci2014}. Modeling the same continuum data,  \citet{Testi2016} found a disk mass of $\sim4.5$\,M$_{\rm{Jup}}$ for a gas-to-dust ratio of 100 and a dust opacity $\kappa_{890\mu \rm{m}}=2\,$cm$^2$/g. The CIDA\,1 disk is detected very clearly in the CO(3-2) line \citep{Ricci2014}, with a velocity gradient consistent with rotation. In the data reported by \cite{Ricci2014}, there is no evidence of   structures either in the dust or gas. 
The SED of CIDA\,1 has been modeled most recently by \cite{Hendler2017}. While the outer radius cannot be constrained by the SED alone, there is no evidence of an inner cavity ($R_{\rm{in}}$=$0.3^{+1.5}_{-0.3}$\,au). The far-infrared fluxes for CIDA\,1 are discussed in \citet{Daemgen2016}; in their sample of very low mass stars, the disk of CIDA\,1 is one of the brightest at 100-200$\,\mu$m.

\section{Observations}		 \label{sec:observations}
CIDA\,1 was observed with ALMA on August 12, 2016 at Band 7 (frequency of about 338 GHz; ALMA project ID: 2015.1.00934.S; PI: L. Ricci). During the observations 38 antennas were available. The total time on-source was about 50 minutes, and observations covered unprojected baselines between about 15 and 1462 m.

The ALMA correlator was configured to record dual polarization
with four separate spectral windows. Two spectral windows were dedicated to the continuum (bandwidth of 2 GHz), and have central frequencies of 332.0 and 344.0 \,GHz, respectively. The mean frequency of these observations 338.0\,GHz, corresponding to a wavelength of about 887\,$\mu$m. Two more spectral windows, with central frequencies of 330.6 and 345.8\,GHz and channel width of 61.0\,kHz, were used to observe CO isotopologues. These data are not discussed in this paper.

\begin{figure*}
 \centering\tabcolsep=0.1cm
 \tabcolsep=0.1cm
 \begin{tabular}{cc}
  	\includegraphics[width=\columnwidth]{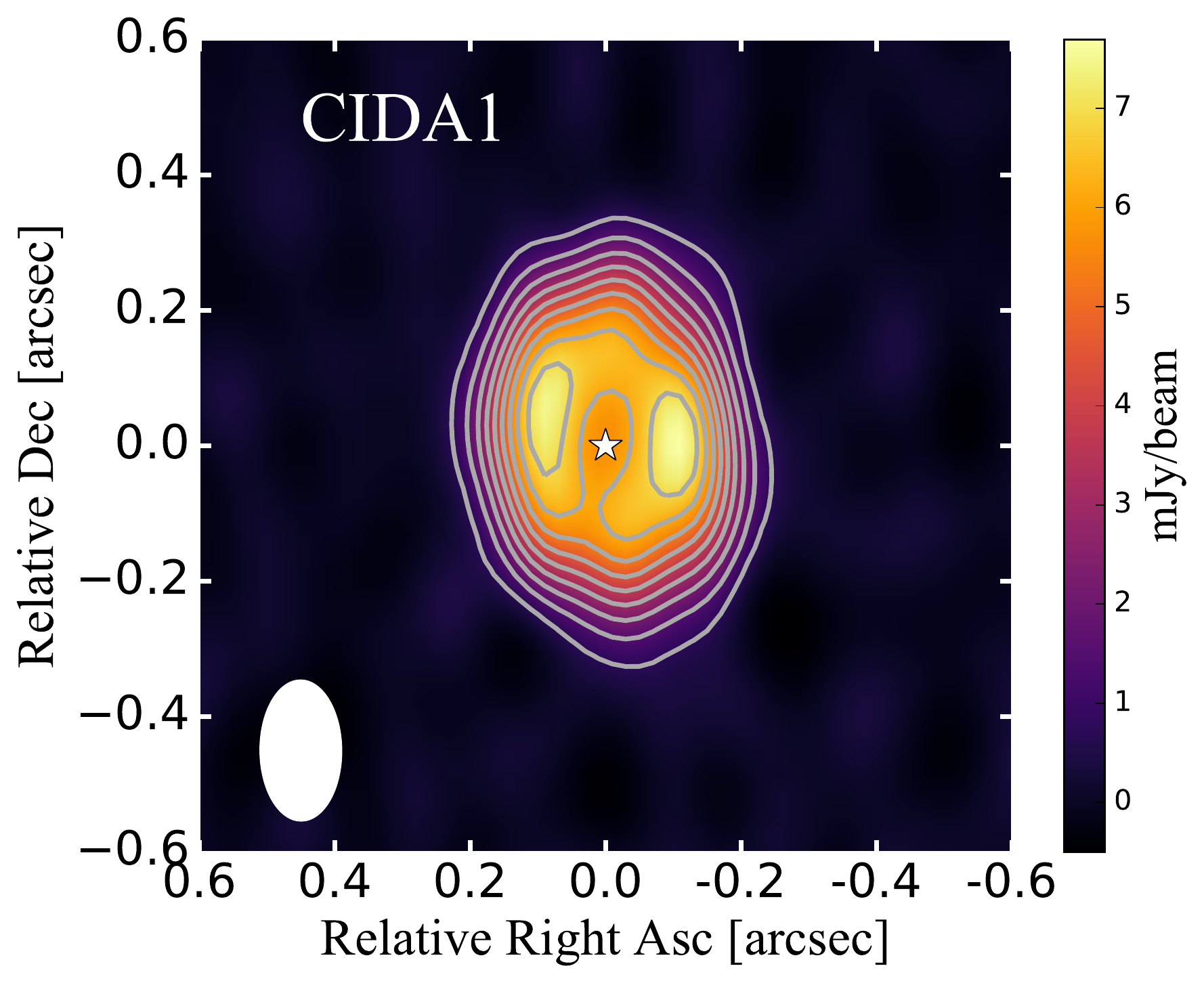}&
	\includegraphics[width=\columnwidth]{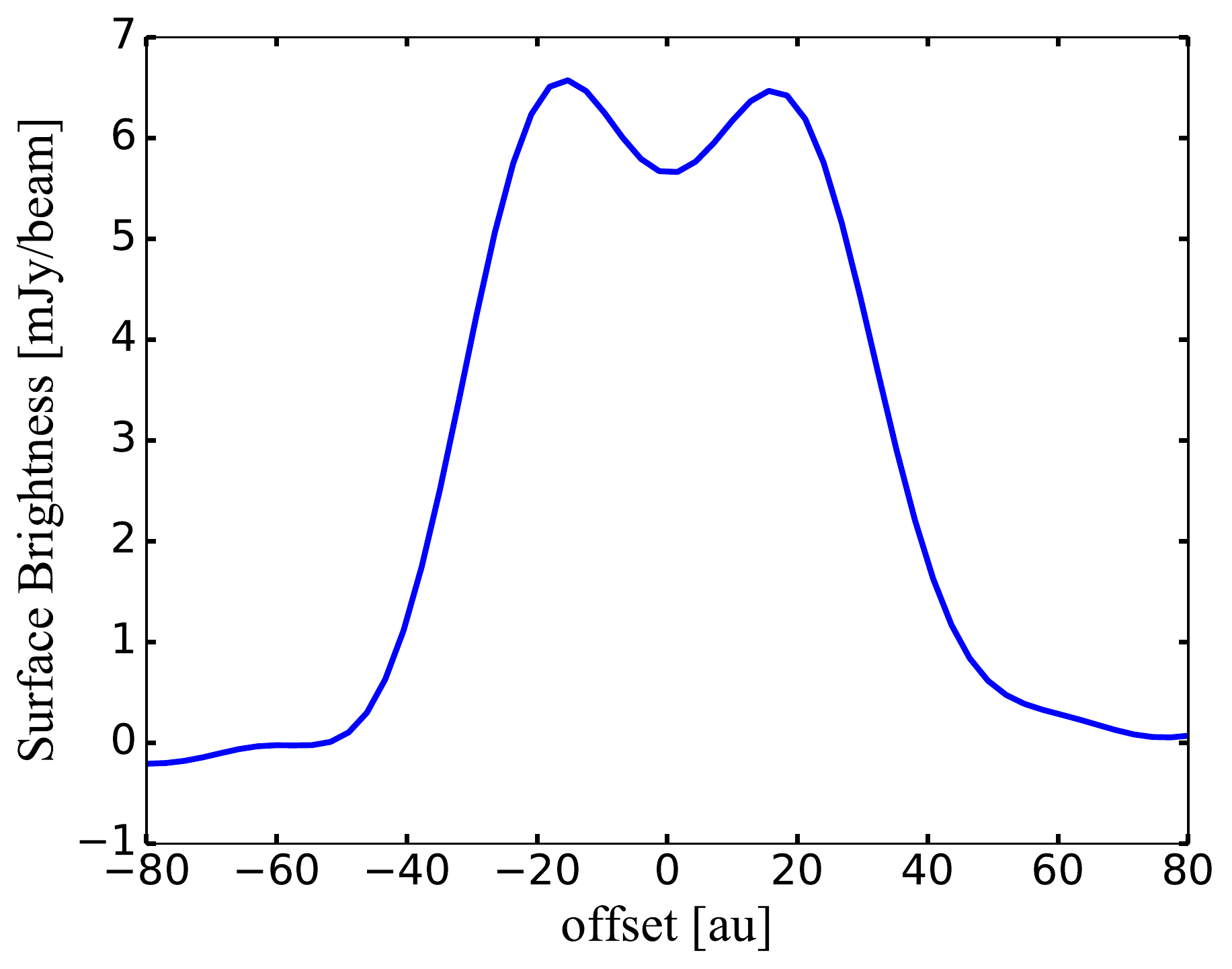}
\end{tabular}
\caption{Left panel: ALMA observations of dust continuum emission for CIDA\,1 at 887$\mu$m, with the contour overlaid at 10,20,...,90\% of the peak ($\sim7.7$\,mJy).  The beam size is indicated at the bottom left corner ($0.\arcsec21\times0.\arcsec12$). Right panel: continuum flux along the PA$=12^\circ$ of the disk, the offset is given in au assuming a distance of 140\,pc.}
   \label{ALMA_obs_CIDA1}
\end{figure*}

The ALMA data were calibrated by National Radio Astronomy Observatory (NRAO) staff using the CASA software package~\citep{mcmullin2007}. Simultaneous
observations of the 183\,GHz water line with the water vapor radiometers were used to reduce atmospheric phase noise before using J0403$+$2600 for standard
complex calibration. The frequency-dependent bandpass and absolute flux scale were calibrated using J0510$+$1800. While no Solar System object was observed as part of our program scheduling block, the flux calibration scale is tied to the Butler-JPL-Horizon 2012 models, resulting in an accuracy of about $10\%$. 

Following the standard ALMA pipeline processing, we performed two rounds of self-calibration on phase only (20s solution intervals) and amplitude and phase (200s solution intervals). The resulting visibilities were time and spectrally averaged, excluding the CO(3-2) frequency range. 

Before imaging and further modeling of the visibilities, the phase center of the data was corrected by an offset of $[0.\arcsec13, -0.\arcsec42]$ using the algorithm \texttt{fixvis} in CASA. This offset was obtained using \texttt{uvmodelfit}, for which we assumed a simple Gaussian or a disk model. Both approaches provided the same results (within the uncertainties) for the offset, position angle (PA), and disk inclination. Both models (Gaussian and disk model) give good convergence, however, as a test we also perform the fit taking only short baselines ($\leq$200 or $\leq$400 k$\lambda$), which do not resolve the disk inner cavity (see below). These tests give similar results as using the entire dataset. Eventually we adopted as disk center position $\alpha_{2000}$=04:14:17.62, $\delta_{2000}$=+28:06:09.28. The PA and disk inclination were also derived from the \texttt{uvmodelfit}, PA=$12.0^\circ\pm0.9^\circ$, and $i=37.3^\circ\pm0.6^\circ$, these values are used for the analysis presented in this paper. 

Imaging was performed with the \texttt{clean} algorithm in CASA using uniform weighting. The resulting image, presented in the left panel of Fig.~\ref{ALMA_obs_CIDA1}, has an rms of 0.15\,mJy\,beam$^{-1}$ and a synthesized beam of $0.\arcsec21\times0.\arcsec12$ (FWHM).

\section{Results} 		\label{sec:results}

The dust continuum emission at 887$\mu$m shows evidence for an inner cavity in the dusty disk (Fig.~\ref{ALMA_obs_CIDA1}, left panel).
The right panel of Fig.~\ref{ALMA_obs_CIDA1} shows the surface brightness along the PA of the disk from the image in the  left panel, revealing ring-like structure peaking at $\sim$20\,au (assuming a distance to the target of 140\,pc) from the central star, and a substantial cavity within the ring, a clear sign of a TD around a very low mass star.

The total flux obtained from the image is $\sim 3.6\times10^{-2}$Jy, with a noise level or rms ($\sigma$) of the observations of 0.2\,mJy. With this flux, we calculate the disk dust mass assuming optically thin emission, as \cite{hildebrand1983}, 

\begin{equation}
	M_{\mathrm{dust}}\simeq\frac{{d^2 F_\nu}}{\kappa_\nu B_\nu (T(r))},
  \label{mm_dust_mass}
\end{equation}

\noindent where $d$ is the distance to the source taken to be 140\,pc, \ $\kappa_\nu$ is the mass absorption coefficient at a given frequency. We assume a frequency-dependent relation given by  $\kappa_\nu=2.3\,$cm$^{2}$\,g$^{-1}\times(\nu/230\,\rm{GHz})^{0.4}$ \citep[][]{andrews2013}. $B_\nu (T_{\rm{dust}})$ is the Planck function for a given dust temperature $T_{\rm{dust}}$. The correct value of $T_{\rm{dust}}$ is not well known. For the often assumed $T_{\rm{dust}}=$ 20\,K \citep[e.g.,][]{ansdell2016} the dust disk mass is $2.7\times10^{-5}\pm 8.7\times10^{-6}\,M_\odot$ ; the prescription $T_{\rm{dust}}\approx25\times(L_\star/L_\odot)^{0.25}$\,K derived by  \cite{andrews2013}, gives $5.2\times10^{-5}\pm 1.7\times10^{-5}\,M_\odot$, while the relation derived by  \cite{vanderplas2016} gives a dust disk mass of $4.4\times10^{-5}\pm1.4\times10^{-5}\,M_\odot$.  The disk mass obtained in \cite{Testi2016} is $4.5\times10^{-5}M_\odot$. For the calculation of the dust disk mass uncertainty, we include 10\% of uncertainty from the flux calibration,  20\,pc for the uncertainty of the distance to Taurus, and 10\% of uncertainty for the dust opacity. In the following, we will adopt the value of $5.2\times10^{-5}\pm 1.7\times10^{-5}\,M_\odot$ or $5.5\times10^{-2}\pm1.8\times10^{-2}\,M_{\rm{Jup}}$ for the dust disk mass, since we will compare the properties of CIDA\,1 to those of the disks in Taurus  analyzed in \cite{andrews2013}.

\section{Data Analysis}  		\label{sec:analysis}

To describe the disk morphology, we performed  the following analysis in the visibility domain. We assume the center, PA, and inclination obtained from \texttt{uvmodelfit} in CASA as described in Sect.~\ref{sec:observations} to deproject and bin the data. We assume an axisymmetric disk  since the image in Fig.~\ref{ALMA_obs_CIDA1} does not show significant asymmetries. Therefore, we only fit the real part of the visibilities. In this case, the Fourier transform of a symmetric brightness distribution can be expressed in terms of the zeroth-order Bessel function of the first kind $J_0$ of the de-projected $uv$-distance, such that

\begin{equation}
V_{\rm{Real}} (r_{uv})=2\pi\int^\infty_0 I(r) J_0(2\pi r_{uv}r)r dr,
\label{eq:real_part}
\end{equation}

\noindent where $r_{uv}=\sqrt{u_\phi^2\cos{i}^2+v_\phi^2}$, being $u_\phi$ and $v_\phi$ the deprojected $uv$-distances. For the intensity profile $I(r)$, we consider a model with a single Gaussian ring. The fitting is conducted using a Markov chain Monte Carlo (MCMC) method with three free parameters ($r_{\rm{peak}}, \sigma_{\rm{width}}$, $F_{\rm{total}}$), such that the intensity radial profile is given by 

\begin{equation}
I(r)=C\exp\left(-\frac{(r-r_{\rm{peak}})^2}{2\sigma_{\rm{width}}^2}\right),
\label{eq:ring_model}
\end{equation}

\noindent where the constant $C$ is related to the total flux of the disk as 

\begin{equation}
C=\frac{F_{\rm{total}}}{\int^\infty_0 I(r) J_0(0)r dr}. 
\end{equation}

To perform the fit, we use {\it emcee} \citep{foreman2013}, which allows us to efficiently sample the parameter space in order to maximize the likelihood result for each model. We follow the procedure as in \cite{pinilla2017a}. The parameter space explored by the Markov chain in this model is:  $r_{\rm{peak}}\in[1, 150]\,\rm{au}$, $\sigma_{\rm{width}}\in [1, 50]\,\rm{au}$, and $F_{\rm{total}}\in[0.0, 0.1]\,\rm{Jy}$.

For the radial grid, we assume $r\in[0-500]\,$au with linearly separated steps of 0.25\,au.  We allow the Markov chain to sample the parameter space for one thousand steps, with two hundred walkers. The maximum autocorrelation time is $\sim100$ for all cases, and we examine the last 100 steps to check the convergence of the fit. Figure~\ref{triangle_plot} shows the results from the MCMC fit for CIDA\,1. The result for the best-fitting parameters is:

\begin{figure}
 \centering
  	\includegraphics[width=\columnwidth]{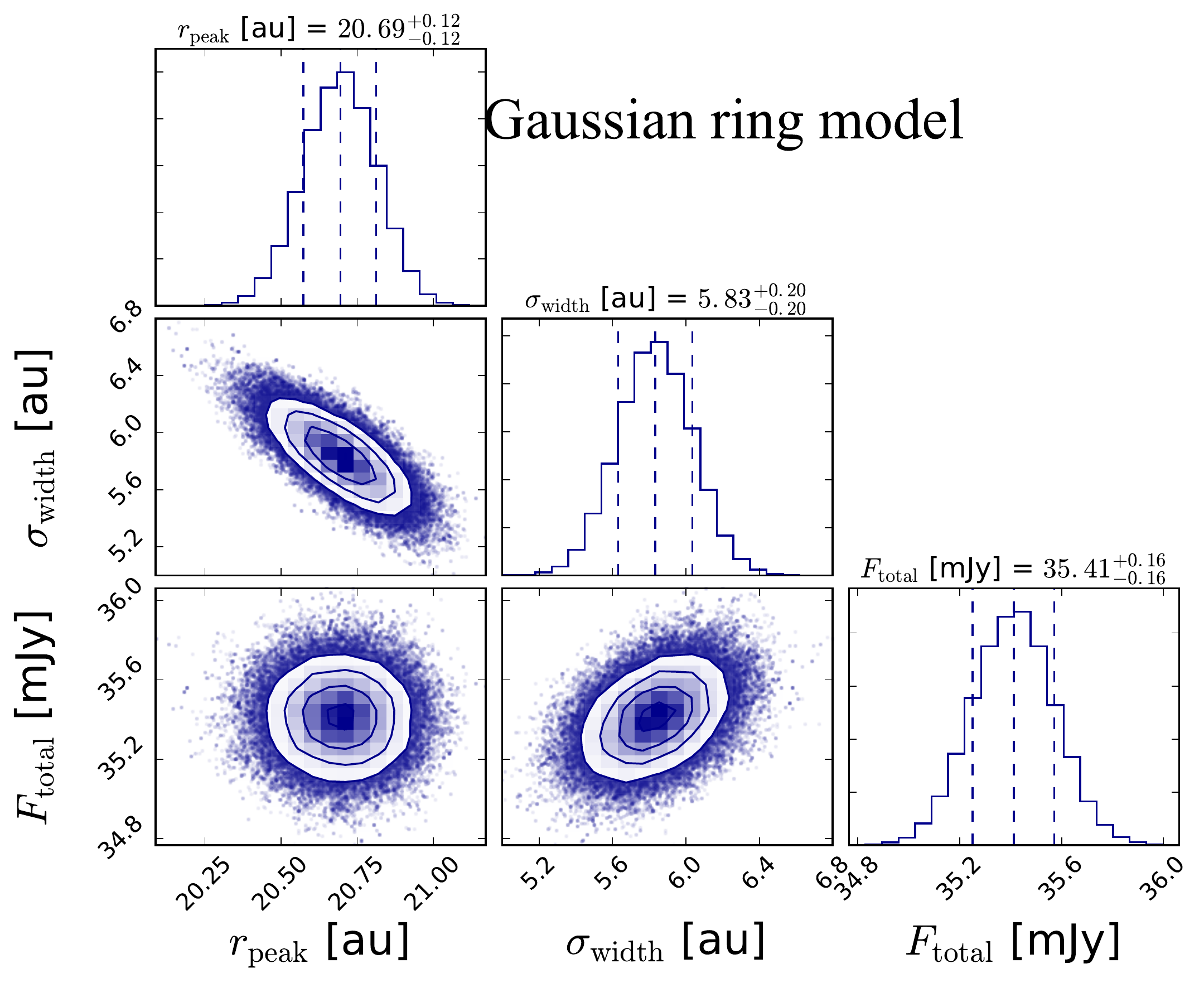}
\caption{Representation of the MCMC results for CIDA\,1, showing the one-dimensional and two-dimensional posterior
distributions for the MCMC fit. The plot shows the posterior sampling provided by the last 700 steps of the 200 walkers chain. The median values and the 1$\sigma$ standard deviation of the best-fitting parameters are shown is vertical dashed lines.}
   \label{triangle_plot}
\end{figure}

\begin{eqnarray}
r_{\rm{peak}}=20.7^{+0.1}_{-0.1} \,\rm{au}\nonumber\\
\sigma_{\rm{width}}=5.8^{+0.2}_{-0.2}\,\rm{au}\nonumber\\
F_{\rm{total}}=35.4^{+0.2}_{-0.2}\,\rm{mJy}.
\label{best_fit_values}
\end{eqnarray}

\begin{figure*}
 \centering
 \begin{tabular}{ccc}
  	\includegraphics[width=6cm]{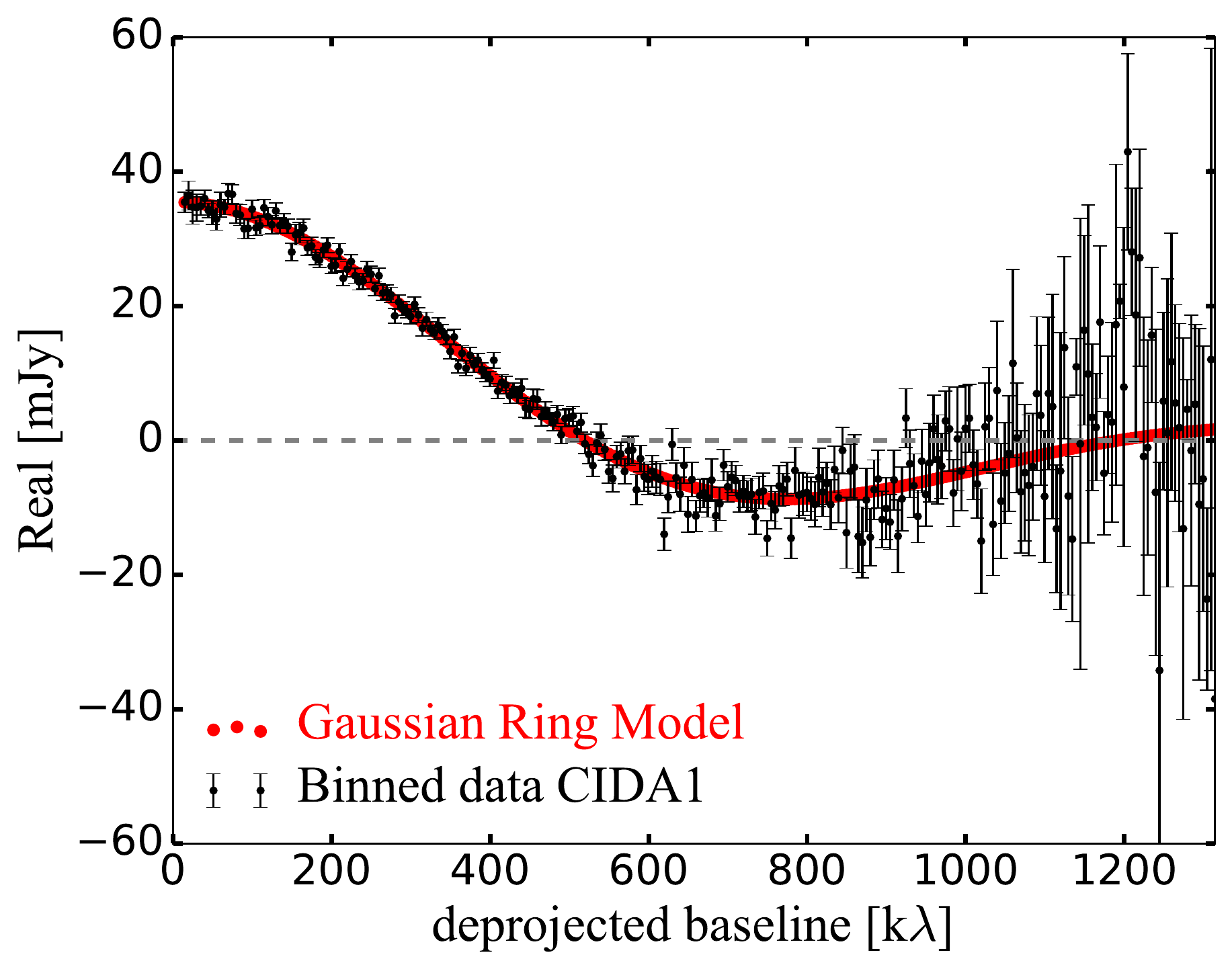}&
	\includegraphics[width=6cm]{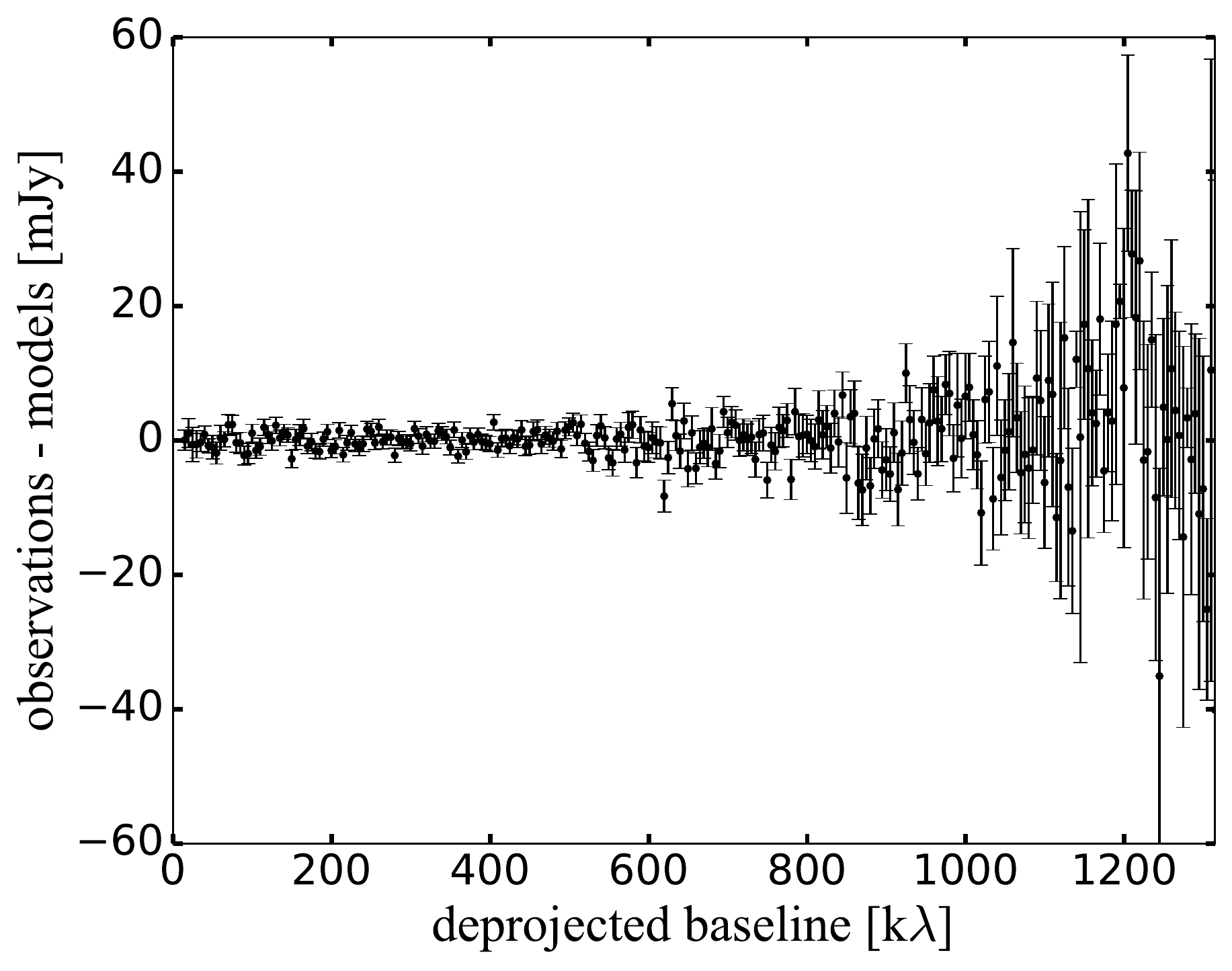}&
	\includegraphics[width=6cm]{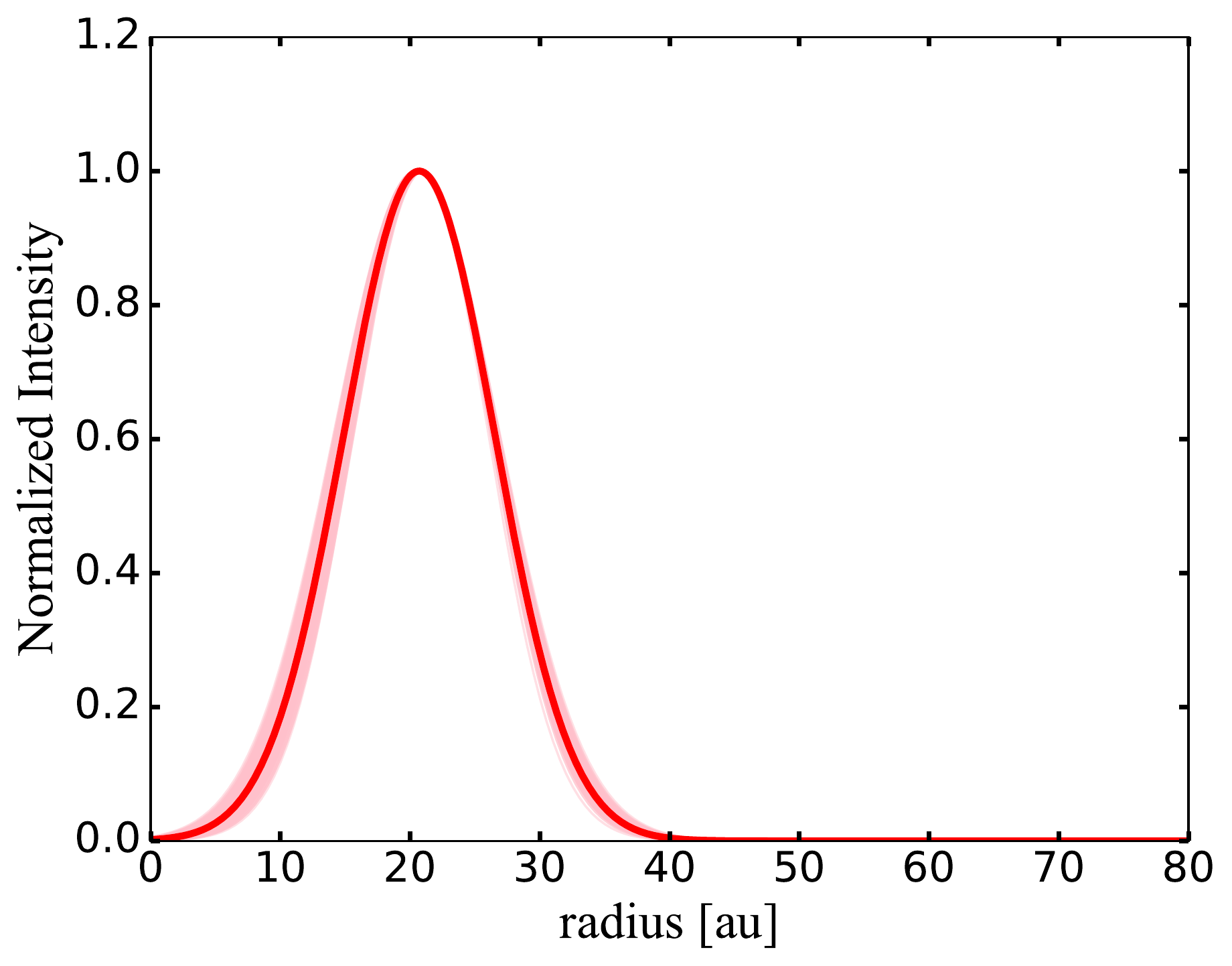}\\
	\includegraphics[width=6cm]{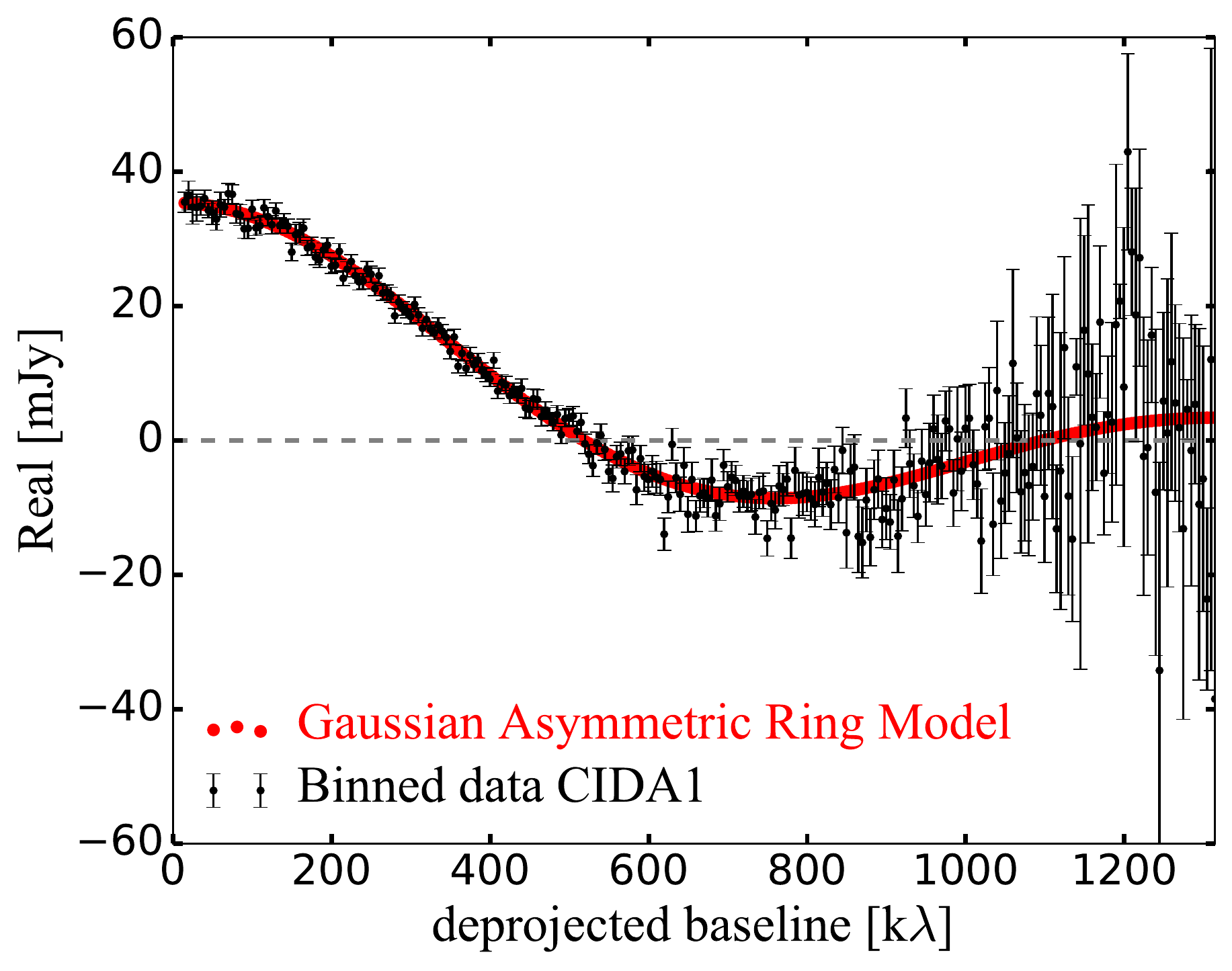}&
	\includegraphics[width=6cm]{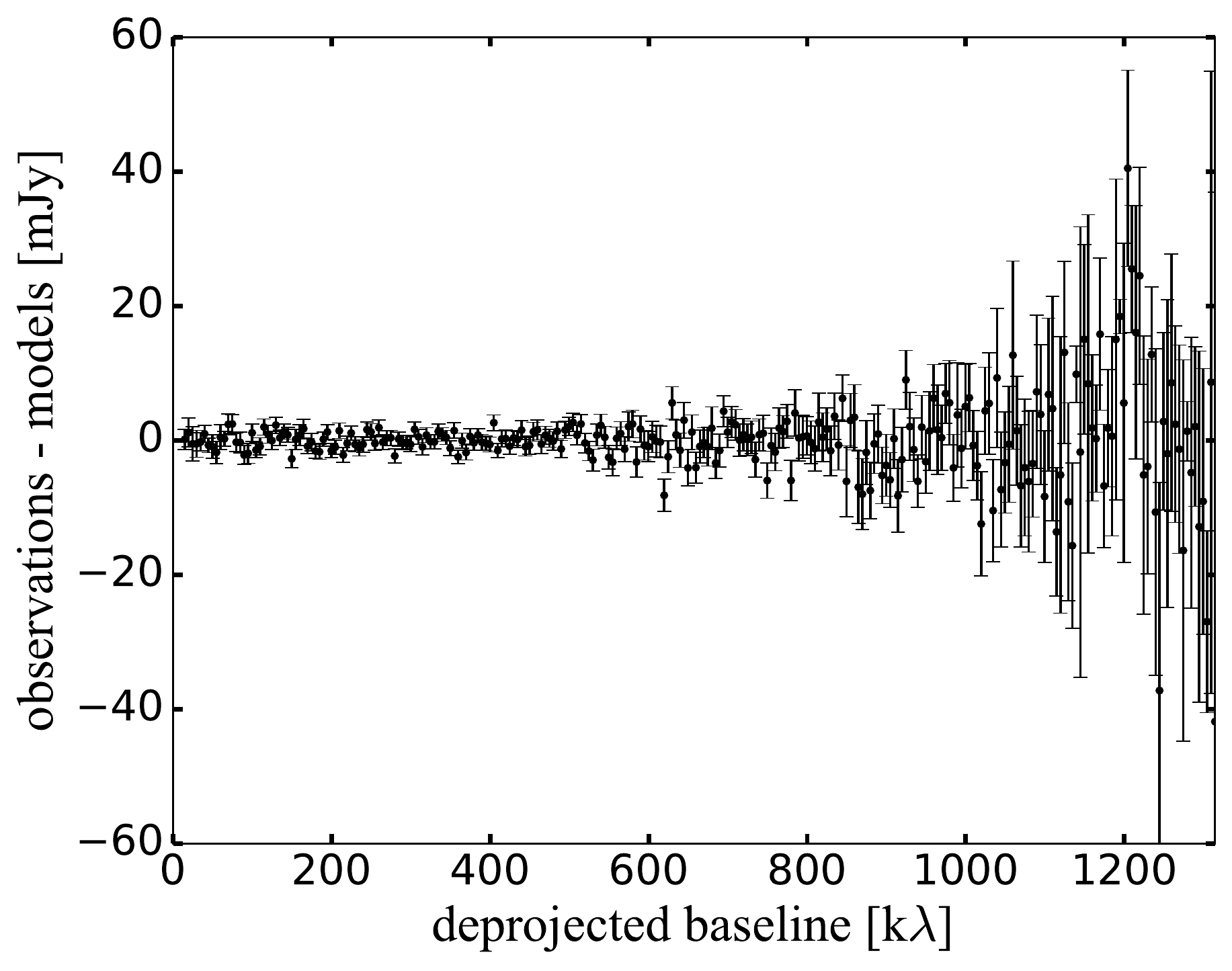}&
	\includegraphics[width=6cm]{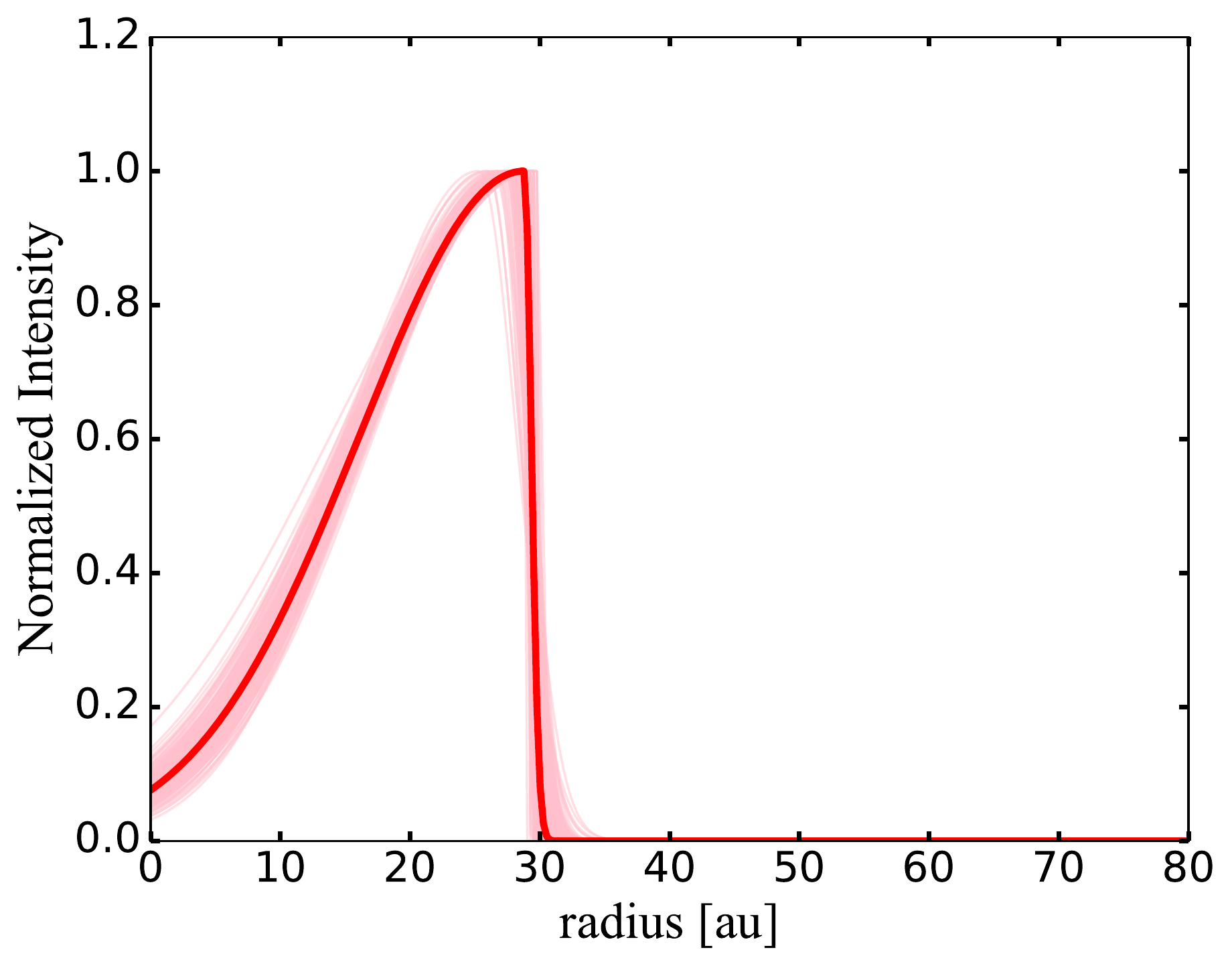}
\end{tabular}
\caption{Left panel: Real part of the visibilities for CIDA\,1. The model with the best-fit parameters is over-plotted.  Middle panel: residuals after subtracting the best-fit model from data. Right panel: Best fit intensity model, which is normalized to the value at the location of peak of the ring. In addition, 200 samples from the chain are over-plotted (light pink lines). The top panels correspond to a model with a radially symmetric ring  (Eq.~\ref{eq:ring_model}) and the bottom panels correspond to a Gaussian with different inner and outer widths with respect to the peak}
   \label{model_vs_observations}
\end{figure*}

Top panels of Fig.~\ref{model_vs_observations} show the real part of the deprojected and binned visibilities for CIDA\,1. The model (Eq.~\ref{eq:ring_model}) with the best-fit parameters is over-plotted (Eq.~\ref{best_fit_values}).  In addition, the residuals and the best-fit model of the intensity profile are also shown. The best-fit model clearly show evidence for a cavity, with radius $\sim20\,$au, defined as the radial position where the peak of the ring of emission is located. 

As an experiment, we also perform a fit assuming an asymmetric Gaussian with different inner and outer widths with respect to the peak (bottom panels of Fig.~\ref{model_vs_observations}). In this case, the intensity profile is given by

\begin{equation}
I(r)=\left\{ \begin{array}{rcl}
C\exp\left(-\frac{(r-r_{\rm{peak}})^2}{2\sigma_{\rm{width, int}}^2}\right) &\mbox{for} & r\leq r_{\rm{peak}}\\
C\exp\left(-\frac{(r-r_{\rm{peak}})^2}{2\sigma_{\rm{width, ext}}^2}\right) &\mbox{for} & r> r_{\rm{peak}},
\end{array}\right.
\label{eq:asymmetric_model}
\end{equation}

\noindent where $\sigma_{\rm{width, int}}$ and $\sigma_{\rm{width, ext}}$ correspond to the inner and outer width of the Gaussian with respect to the peak.  

This experiment is motivated by the expected intensity profiles due to particle trapping in a single pressure bump \citep{pinilla2017a},  which show that the ring-like emission at early times of the evolution ($\sim1$\,Myr) has a sharper rise and a more extended outer tail than at  later times ($\sim5$\,Myr), when the ring is more radially symmetric. This is a result of particle growth and drift from the outer part toward the pressure maximum. The parameter space explored in this case is  $r_{\rm{peak}}\in[1, 150]\,\rm{au}$, $\sigma_{\rm{width, int}}\in [1, 50]\,\rm{au}$, $\sigma_{\rm{width, ext}}\in [1, 50]\,\rm{au}$, and $F_{\rm{total}}\in[0.0, 0.1]\,\rm{Jy}$. The best-fitting parameters in this case are: 

\begin{eqnarray}
r_{\rm{peak}}=28.8^{+0.5}_{-0.9} \,\rm{au}\nonumber\\
\sigma_{\rm{width, int}}=12.7^{+0.8}_{-0.8}\,\rm{au}\nonumber\\
\sigma_{\rm{width, ext}}=0.6^{+0.7}_{-0.4}\,\rm{au}\nonumber\\
F_{\rm{total}}=35.4^{+0.2}_{-0.2}\,\rm{mJy}.
\label{best_fit_values}
\end{eqnarray}

To quantify which of the two models (radially symmetric ring or radially asymmetric ring) provides a better fit, we obtain the Bayesian Information Criterion (BIC), which is calculated {as $\rm{BIC}= \ln(N)N_{\rm{variables}}-2\ln(\hat{L})$, being $N$ the number of data points, $N_{\rm{variables}}$ the number of variable parameters, and $\hat{L}$ the maximum likelihood value. The model with the lowest BIC value is the radially asymmetric ring model, and the difference between the two BICs is 11.7, which implies a strong evidence against the radially symmetric model.

In both cases the ring is quite narrow and the disk  is very compact (Fig.~\ref{model_vs_observations}). In particular, the radially asymmetric ring model shows that the outer tail decreases sharper than the inner width of the ring,  suggesting a very effective drift of particles from the outer disk. \cite{pinilla2018} analyzed a total of  29\,TDs observed with ALMA and found that less than 10\% of the disks in the sample (including CIDA\,1) have a much sharper tail compared to the inner width of the ring. These few targets, however, do not have any common property (they  have different stellar masses, disk masses, or accretion rate), leaving us an open question why in few cases radial drift is more efficient in the outer disk.

The outer radius of the dust continuum emission derived from our analysis of the ALMA Cycle\,3 observations is more compact ($\sim 30$\,au) than previously inferred from ALMA Cycle\,0 data \citep{Ricci2014, Testi2016}. The uv-coverage from the Cycle\,0 data is up to $\sim350\,\rm{k}\lambda$, which is shorter than the position of the null in the visibilities ($\sim500\,\rm{k}\lambda$) that evidences the existence of a cavity. The most current data, covers a factor of almost four more than Cycle\,0 data (up to $\sim1350\,\rm{k}\lambda$). Because the Cycle\,0 had a very short u-v coverage and the disk was marginally resolved, the disk size could be over-estimated in previous analysis, in particular if the fit was forced to be a power-law or a tapered profile, as in \cite{Ricci2014, Testi2016}. 

\section{Discussion} 		\label{sec:discussion}

When compared with a sample of T-Tauri stars in Taurus, CIDA\,1 has a rather massive disk  for its stellar mass, as shown in the left panel of Fig.~\ref{macc_mdust_mstar}.  For this figure, we considered the stellar masses calculated using the evolutionary models of \cite{Baraffe1998} and the millimeter fluxes and dust masses obtained in \cite{andrews2013}. Moreover, we include \citep[or update if they were in the previous sample from][]{andrews2013} the recent targets reported by \cite{ward2018}. To calculate the dust disk mass, we follow the same procedure as in \cite{andrews2013}, and we use Eq.~\ref{mm_dust_mass} assuming for all targets $\kappa_\nu=2.3\,$cm$^{2}$\,g$^{-1}\times(\nu/230\,\rm{GHz})^{0.4}$ and $T_{\rm{dust}}\approx25\times(L_\star/L_\odot)^{0.25}$\,K.

\begin{figure*}
 \centering
 \tabcolsep=0.1cm 
   \begin{tabular}{cc}   
   	\includegraphics[width=\columnwidth]{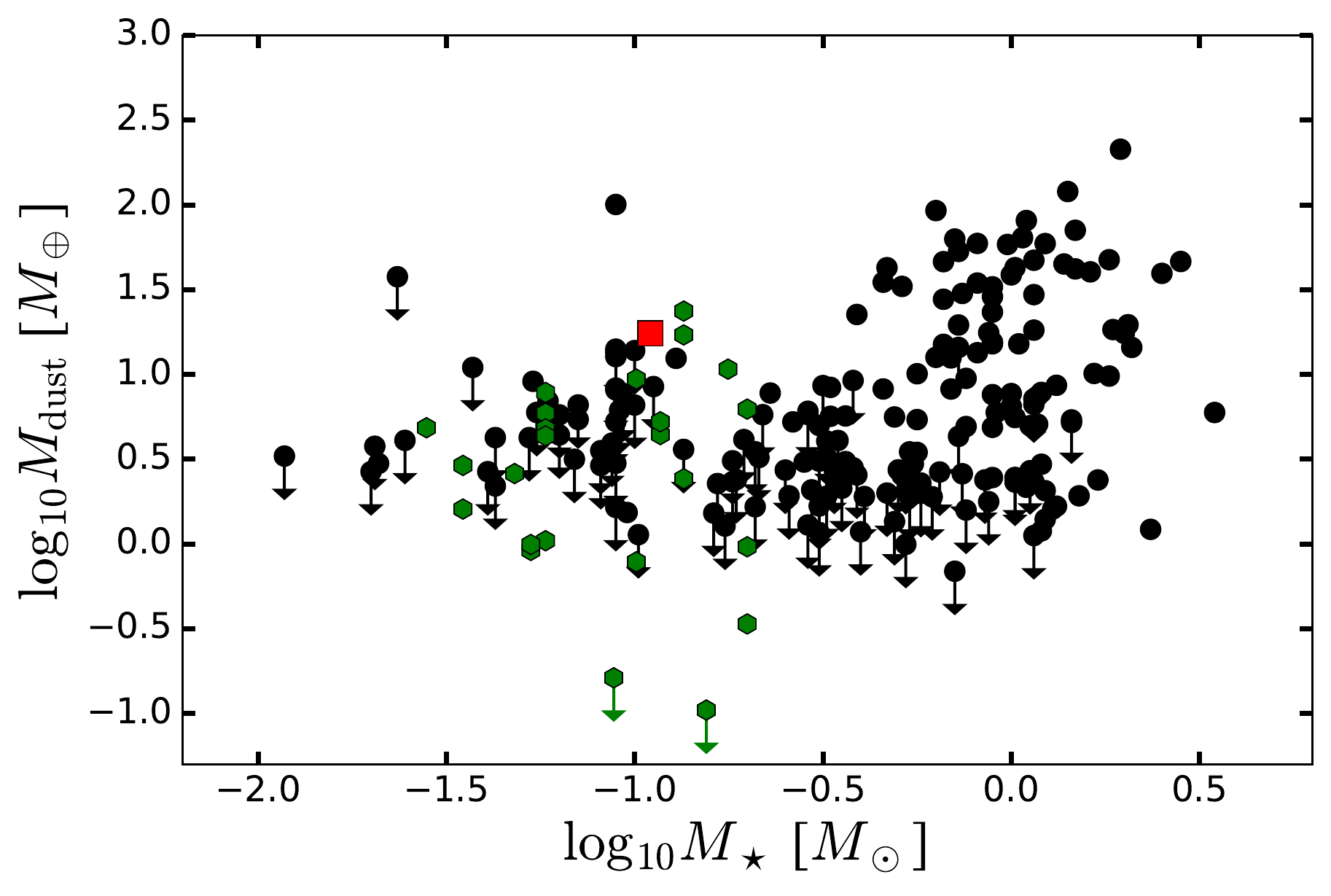}&\includegraphics[width=\columnwidth]{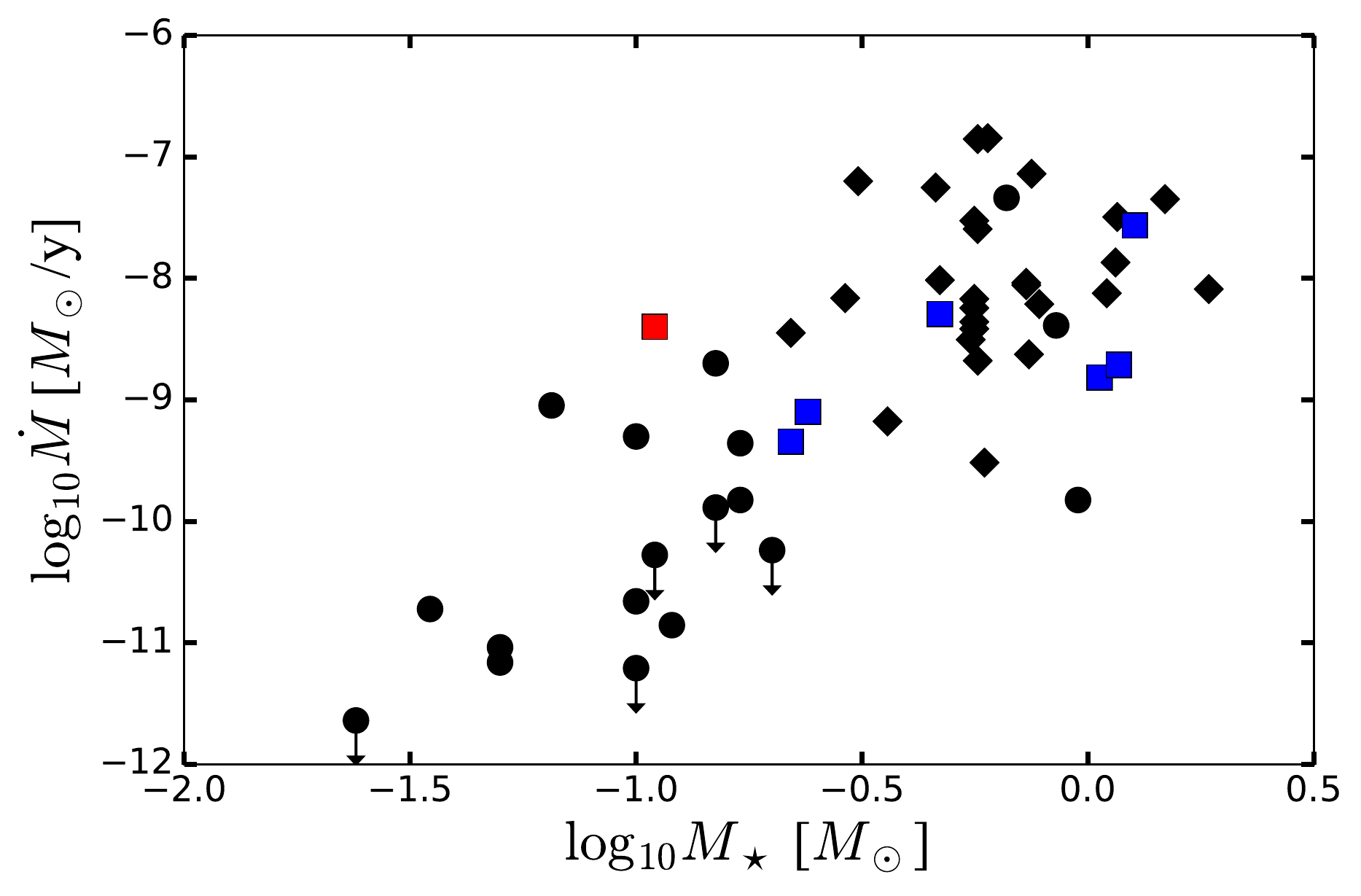}
   \end{tabular}
    \caption{Left panel: Dust mass versus stellar mass for Taurus Class II objects. Dots and upper limits from \cite{andrews2013}, assuming a  dust temperature as $T_{\rm{dust}}\approx25\times(L_\star/L_\odot)^{0.25}$\,K, and the evolutionary models from \cite{Baraffe1998} to calculate the stellar mass. The hexagonal points correspond to the targets reported by \cite{ward2018}, assuming the same procedure as in \cite{andrews2013} to calculate the dust disk mass. The position of CIDA\,1 is shown by the red square.  Right panel: Mass accretion rate versus stellar mass for Taurus Class II objects. Dots from \cite{Herczeg2008}, diamonds from the compilation of \cite{Rigliato2015}. The location of CIDA\,1 is shown by the red square. Known TDs as classified by \cite{Rigliato2015} are identified by blue squares.}
   \label{macc_mdust_mstar}
\end{figure*}

CIDA~1 has a  massive disk for its stellar mass. With a disk mass of  5.5$\pm$1.8\,M$_{\rm{Jup}}$ (assuming a dust-to-gas ratio of 1/100) and a stellar mass of 0.1\,M$_\odot\pm0.15\,$dex, it is $M_{\rm{disk}}/M_{\star} \sim 4.8\times10^{-2} \pm 1.6\times10^{-2}$.  \cite{andrews2013} estimate that less than $\sim$2-5\% of Taurus Class II have a ratio larger than that. The mass accretion rate of CIDA\,1 is also very high, as shown in the right panel of Fig.~\ref{macc_mdust_mstar}. However, it should be kept in mind that the sample of Taurus T-Tauri stars with homogeneously measured mass accretion rates is very small. The ratio $M_{\rm{disk}}/\dot{M}$ is 1.4\,Myr. When compared with other disks in the Lupus and Chamaeleon I star forming regions \citep{Manara2016b,Mulders2017} this ratio is in the lower $\sim$5\% of the ratios measured in full disks, and it is the lowest for transition disks with large cavities \citep{najita2015, Manara2016b}. This makes this target an exceptional one.

The cavities of TDs can have different origins, including internal photoevaporation from the stellar irradiation \citep[e.g.,][]{alexander2007, owen2012}, planet disk interaction \citep[e.g.,][]{rice2006, zhu2011, pinilla2012}, dead-zones \citep[e.g.,][]{regaly2012, flock2015, pinilla2016}, or a combination of any of these phenomena \citep[e.g.,][]{rosotti2013}. 

\paragraph{Photoevaporation} Models of photoevaporation predict small cavities ($\lesssim$10\,au) and low accretion rates ($\lesssim10^{-9}\,M_\odot\,$year$^{-1}$) \citep[see e.g., Fig.~2 from][]{owen2017}. The cavity size inferred from the ALMA observations of CIDA\,1  and its mass accretion rate  ($\dot M\sim 4 \times 10^{-9} \,M_\odot\,$year$^{-1}$) are very large and fell outside the range predicted by these  models.  Recently, \citet{ercolano2018} showed that  X-ray photoevaporation  in disks with modest gas-phase depletion of carbon and oxygen can explain TDs with a large diversity of accretion rates and cavity sizes.  However, it is important to note that the models in \citet{ercolano2018} and \citet{owen2017} are obtained assuming a central star with stellar mass of 0.7\,M$_\odot$,  much more massive than CIDA\,1. Although more appropriate models are clearly needed, for the moment photoevaporation does not appear to be the most likely explanation of the CIDA~1 inner cavity. 

\paragraph{Dead zones}
Dead zones are a potential origin for TD-like structures. Dead zones are formed because X-rays and far-UV rays from the central star cannot penetrate the deep and dense disk midplane, which leads to a low-ionized region where MRI is suppressed. The size and shape of a dead zone depends on different factors such as chemistry and dust properties \citep{Dzyurkevich2013}. In particular, the extension of the dead zone depends on the disk density. \citet{pinilla2016} implemented changes in the effective $\alpha-$viscosity \citep{shakura1973} that depend on the disk gas surface density. They showed that dead zones can open gaps/cavities and reproduce structures as seen in TDs. According to MHD simulations, the typical values of the gas surface density threshold for which X-rays and far-UV cannot penetrate the disk is ~10-15\,g/cm$^2$. In the case of low mass disks, these values are only reached in the inner disk. To calculate where this threshold is reached in the particular case of CIDA\,1, we assume that the initial gas surface density (before the cavity was opened) is an exponentially tapered profile \citep[e.g.,][]{Guilloteau2011}, that is $\Sigma(r)=\Sigma_0(r/r_0)^{-\gamma} \exp[-(r/r_c)^{2-\gamma}]$, with $r_0=100$\,au, $r_c=80$\,au, $\gamma=1.5$, and we take the dust mass from the current ALMA observations and assume a dust-to-gas ratio of 1/100 to calculate the normalization factor $\Sigma_0$. The values of $r_c$ and $r_0$ are not constrain from current gas observations of  CIDA\,1, but we assume that the gas density is around $\sim2.6$ times more extended than the millimeter dust, as observed in different protoplanetary disks \citep[e.g.,][]{Gregorio2013, Walsh2014}. Under this assumption, the gas surface density threshold for which X-rays and far-UV cannot penetrate the disk is reached at $\sim$5au, preventing the formation of a extended dead zone and hence of a very large cavity. 

If this threshold is lower ($\sim2$g/cm$^2$), the outer edge of the dead zones can be moved outward to $\sim$8-10au. In this case, however, the gas surface density bump that is formed at $\sim10\,$au dissipates on a very short timescale ($\lesssim$0.2\,Myr) due to viscous evolution. Hence, forming a long-lived large cavity due to a dead zone in a very low mass disk is unlikely. 

\paragraph{Embedded Planets} A massive planet can open a gap in the gas surface density and create an empty cavity in millimeter- and centimeter-sized grains. We calculate the minimum mass planet to open a gas gap under the CIDA\,1 disk conditions \citep{crida2006, pinilla2017b}, assuming an $\alpha-$viscosity value of $10^{-4}$, and obtain values for the planet-star mass ratio of  $\sim 2.5\times10^{-3}$, which corresponds to around $1.0\times M_{\rm{Saturn}}$ by assuming $M_\star=0.1\,M_\odot$, and the planet located at 15\,au in order to have the cavity at $\sim20\,$au (Fig.~\ref{Crida_criterion}). Taking the dust disk masses calculated in Sect~\ref{sec:results}, and assuming a gas-to-dust ratio of 100, the disk mass is between $\sim10\,M_{\rm{Saturn}}$ to $\sim18\,M_{\rm{Saturn}}$. These estimates show that the presence of such a planet is a realistic scenario to explain the disk morphology of CIDA\,1. When assuming an intermediate $\alpha-$viscosity value of $10^{-3}$,the minimum mass planet to open a gap in the gas surface density increases, and the planet-star mass ratio is $\sim 4.0\times10^{-3}$, which corresponds to a  $\sim14\,M_{\rm{Saturn}}$ planet. Assuming that CIDA\,1 is 1\,Myr old and its accretion rate is $4\times 10^{-9}$ M$_\odot$/yr implies that at least $\sim44\%$ of the total initial disk mass was used for the formation of such a planet. This suggests that an embedded massive planet is a plausible explanation for the cavity formation only for a relatively low viscosity disk.

\begin{figure}
 \centering
  	\includegraphics[width=\columnwidth]{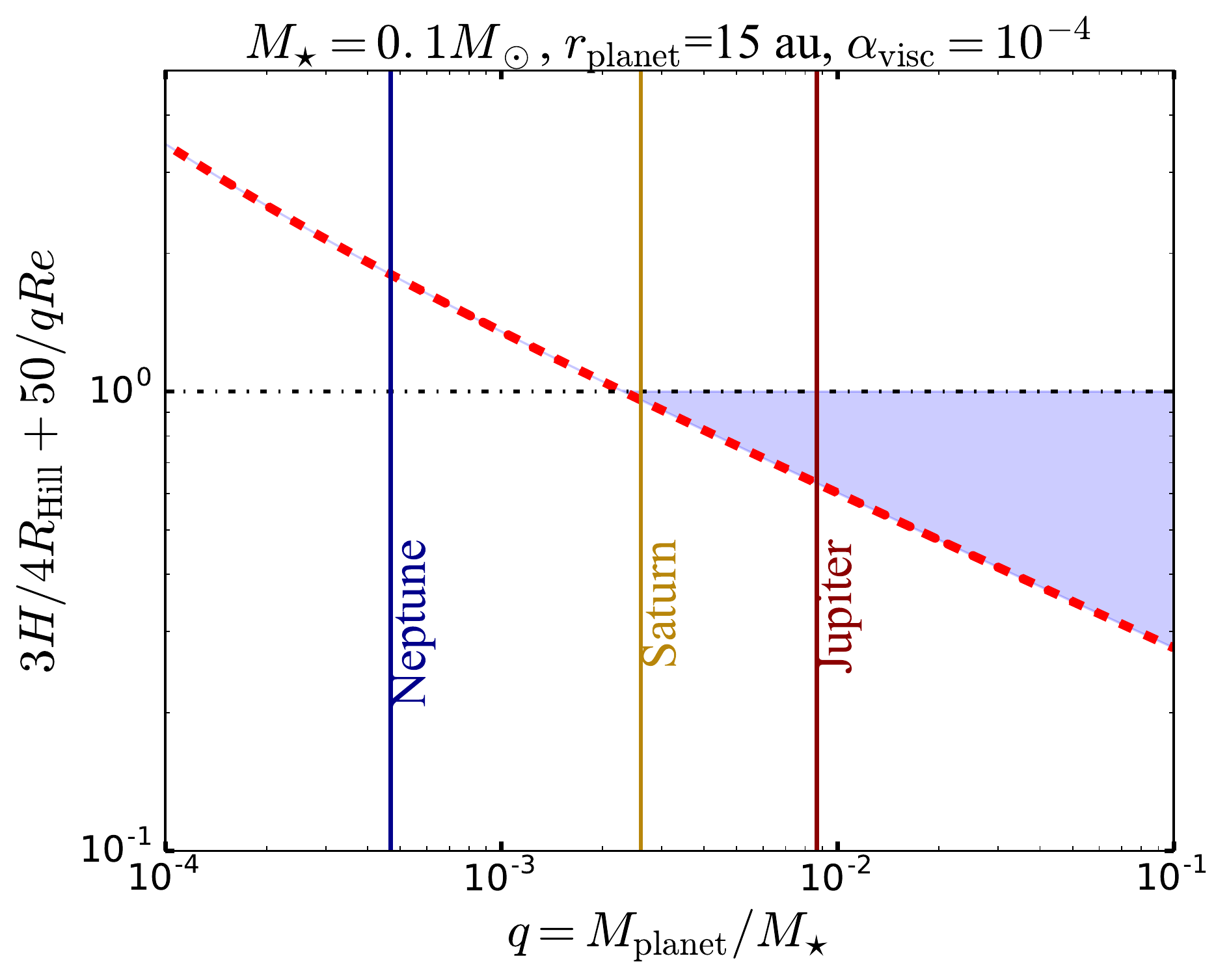}
\caption{Criterion to open a gap in the gas surface density \citep{crida2006} assuming the disk parameters for CIDA\,1. When the criterion (in the y-axes) is lower than unity, a planet of such mass can open a gap in the gas surface density of the disk, generating particle trapping and an empty dust cavity.}
   \label{Crida_criterion}
\end{figure}

\paragraph{Circumbinary disk} 
It is possible that instead of hosting a massive planet, CIDA\,1 is a binary system, with a companion that is responsible for the cavity formation. In this case, the cavity shape depend on the mass ratio, eccentricity, and disk viscosity \citep[e.g.,][]{Artymowicz1994, miranda2017}. The pressure maximum in the binary case can be as far as four times the separation of the binary system. This means that a binary separation of 5\,au is enough to explain the cavity at 20\,au \citep[see Fig.~2 in][]{miranda2017}. To our knowledge, there is no evidence for the presence of a companion to CIDA-1 \citep{Kraus2012}, but separations of a few au have not been probed yet. This scenario would also imply a larger mass of the central source(s), which would make the ratio $M_{\rm{disk}}/M_{\star}$ lower, more in line with the values observed in \cite{andrews2013}.

The accretion rate can be enhanced due to the presence of the companion. The accretion rate is expected to be similar for both stars when there are equal-mass circular binaries. The accretion rate for such systems can be on the order of 10$^{-7}\, M_\odot$/y - 10$^{-8}\, M_\odot$/y \citep{munoz2016}. If the high mass accretion rate is confirmed for CIDA\,1, a tight equal-mass binary system is a plausible scenario. The accretion rate is expected to be variable with a period which depends mostly on the eccentricity. In the case of circular binaries the variability can be five times the binary period, whereas if the binary is eccentric the period is approximately one binary period. Long-term variability of the accretion rate is therefore expected in CIDA\,1 if a binary is responsible for the cavity formation.

\section{Conclusions} \label{sec:conclusions}

We obtained ALMA Cycle\,3 observations of the disk around CIDA\,1 - a very low mass star. These observations reveal a large dust cavity ($\sim$20\,au) characteristic of TDs, making CIDA\,1 one of the few very low mass stars where a cavity has been resolved in the millimeter emission. CIDA\,1 hosts a massive disk for its stellar mass in comparison with other T-Tauri stars in Taurus. In addition, CIDA\,1 has a very low value of  $\dot M/M_{\rm{disk}}$, contrary to most of the TDs observed so far.
 
We modeled the morphology of the dust distribution in the visibility plane and found that the ring like structure around CIDA\,1 is rather narrow, in agreement with models of particle trapping  when radial drift is very efficient, as expected around very low mass stars.  We discussed the potential origins for the cavity formation, including photoevaporation, dead-zones, embedded planets, and a close binary. Due to the large cavity and the high accretion rate of CIDA\,1, photoevaporation and dead zones are unlikely explanations. A massive embedded planet (or multiple planets whose gaps overlap) or a close binary remain as possible explanations for the origin of the cavity. The high accretion rate observed for CIDA\,1 can be explained in  the case of a binary system, where a long-term variability (one to five times the binary period) is expected for the accretion rate.

The existence of large dust-depleted inner regions in disks around low mass stars is of great importance for our understanding the TD phenomenon, as the disk mass and temperature are significantly lower than those of disks around more massive stars. The case of a gap produced by a massive planet, if confirmed, test the universality of the cavity formation processes, and indicates not only that relatively massive planets could form around stars of very different mass, but also that their disk evolution may be similar. This opens the question of how the radial drift barrier is overcome in the first place and if the mechanisms for the formation of the first planetesimals are similar in disks around a large range of stellar masses. Although, the first pressure traps may originate from for example dead zones or zonal flows that arise from magnetohydrodynamical process, it is an unresolved question if these can form in disks around low mass stars or even in the extreme case of circumplanetary disks \citep{zhu2017}, and if they will be sufficiently strong and long-lived to stop the very high radial drift of particles expected in their disks. 

\section*{Acknowledgements}
We are thankful to the referee for the constructive report. P.P. acknowledges support by NASA through Hubble Fellowship grant HST-HF2-51380.001-A awarded by the Space Telescope Science Institute, which is operated by the Association of Universities for Research in Astronomy, Inc., for NASA, under contract NAS 5- 26555. This work was partly supported by the Italian Ministero dell\'\,Istruzione, Universit\`a e Ricerca through the grant Progetti Premiali 2012 -- iALMA (CUP C52I13000140001), by the Deutsche Forschungs-Gemeinschaft (DFG, German Research Foundation) - Ref no. FOR 2634/1 TE 1024/1-1, by the DFG cluster of excellence Origin and Structure of the Universe (\href{http://www.universe-cluster.de}{www.universe-cluster.de}) and by the ERC grant 743029 EASY. C.F.M. acknowledges support through the ESO Fellowship. This paper makes use of the following ALMA data: ADS/JAO.ALMA\#2015.1.00934.S. ALMA is a partnership of ESO (representing its member states), NSF (USA) and NINS (Japan), together with NRC (Canada) and NSC and ASIAA (Taiwan) and KASI (Republic of Korea), in cooperation with the Republic of Chile. The Joint ALMA Observatory is operated by ESO, AUI/NRAO and NAOJ.

\end{document}